\documentclass[prb,twocolumn,superscriptaddress,showpacs,amsmath,amssymb]{revtex4}

\usepackage{graphicx}
\usepackage{latexsym}
\usepackage{amsmath}
\usepackage{amssymb}
\usepackage{amsfonts}
\usepackage{color}
\usepackage{bm}
\usepackage{verbatim}
\usepackage{color}

\begin{document}
\title{Conductance enhancement due to the resonant tunneling into the subgap vortex core states
in normal metal/superconductor ballistic junctions.}

\author{M.~A.~Silaev}
\affiliation{Institute for
Physics of Microstructures, Russian Academy of Sciences, 603950
Nizhny Novgorod, GSP-105, Russia}

\date{\today}

\begin{abstract}
 We investigate the low-energy quantum
transport in ballistic normal metal-insulator -superconductor
junction exposed to a magnetic field creating Abrikosov vortices
in the superconducting region. Within the Bogolubov- de Gennes
theory we show that the presence of the subgap quasiparticle
states localized within the vortex cores near the junction
interface leads to the strong resonant enhancement of Andreev
reflection probability, and the normal-to supercurrent conversion.
The corresponding increase of charge conductance is determined by
the distance from the vortex chain to the junction interface,
which can be controlled by the applied magnetic field.
\end{abstract}


 \maketitle

\section{Introduction}

Recently, the investigation of transport properties of normal
metal- superconducting (N/S) hybrid structures has attracted a
considerable interest. In the classical work by Blonder, Tinkham
and Klapwijk\cite{BKT} it was shown, that at the energies below
the superconducting gap $\Delta_0$ the charge transport can only
be realized via the Andreev reflection at the N/S interface. This
is a two-particle process in which the electrons with low energies
$\varepsilon<\Delta_0$ incident from the normal metal (N)  are
reflected at the N/S interface as holes traversing the backward
trajectories (and vice versa). In accordance with the charge
conservation law, at the superconducting region (S) the Cooper
pairs are formed and the normal current converts into the
supercurrent. For the perfectly transparent N/S interface the
charge doubling due to Andreev reflection results in enhancement
of the subgap conductance by a factor of two compared with the
corresponding normal state conductance. Being the two-particle
process Andreev reflection is strongly suppressed due to QP
scattering at the layer of insulator separating the N and S
regions. Indeed, in case of the small interfactial barrier
transparency $T\ll 1$, the Andreev reflection probability, and
therefore the conductance is proportional to $T^2$ which is
smaller by the factor $T$ as compared to the single- electron
case.

The phenomenon of subgap conductance suppression
 results in the remarkable dependence of transport properties of N/S structures on the
 spatial distribution and symmetry of the superconducting gap.
For example, recently the charge current measurements were
utilized for the direct observation of multi vortex structures in
mesoscopic superconductors \cite{Peeters}. Mixed state of
mesoscopic superconductors is formed by a small number of vortices
and reveals a rich variety of exotic vortex configurations, such
as vortex molecules and multiquantum giant vortices, realizing in
such samples due to the quantum confinement of the Cooper pairs
motion~\cite{Mesovortices}. In Ref.\onlinecite{Peeters} the phase
transitions between different vortex configurations were observed
by the multiple-small-tunnel-junction method, in which several
small tunnel junctions were attached to the mesoscopic
superconductor to simultaneously measure the charge transport at
the different points of the sample. The measured transport
characteristics were related to the local density of states (DOS),
depending on the local supercurrent density and hence on the
configuration of the vortex system. Generally, when there is a
uniform supercurrent flowing at the superconductor, the excitation
spectrum acquires a Doppler shift by the value
${\bf{v}}_s{\bf{p_F}}$, where ${\bf{v}}_s$ is a superfluid
velocity, ${\bf{p}}_F$ is a Fermi momentum. In this case, the
minimal excitation energy is given by
$E_{min}=\Delta_0-{\bf{v}}_s{\bf{p_F}}$. At large distances from
the vortex core ($r\gg\xi$, where $\xi$ is a coherence length) the
Doppler shift model gives quite a good approximation of the
quasiparticle (QP) spectrum with $v_s=\hbar/2mr$ (see
Ref.[\onlinecite{Tinkham}]). As long as the superfluid velocity is
small compared to the critical value $\Delta_0/p_F,$ the Doppler
shift of the gap edge results in the reduction of the height and
broadening of the superconducting DOS peak \cite{esteve}. Close to
the vortex core ($r\sim\xi$) where $v_s\sim \Delta_0/p_F,$ the gap
edge $E_{min}$ goes to zero. However, in this case the Doppler
shift model does not hold: it completely misses one of the
remarkable features of the vortex state: the presence of
low-energy QP states localized within the vortex core. These
vortex core states were found in the pioneering work by Caroli- de
Gennes and Matricon (CdGM)\cite{CdGM} within the more rigorous
approach based on the quasiclassical solution of Bogoliubov- de
Gennes (BdG) equations. It was shown that QP states
 with energy lower than the bulk superconducting gap value
 $\Delta_0$ are localized within the vortex core and have the
discrete spectrum $\varepsilon_\mu$ as a function of the quantized
(half--integer) angular momentum $\mu$. This spectrum of the CdGM
states varies from $\Delta_0$ to $-\Delta_0$ as $\mu$ changes from
$-\infty$ to $+\infty$, crossing zero when $\mu$ changes its sign.
At small energies $|\varepsilon|\ll\Delta_0$ the spectrum is given
by $\varepsilon_\mu\approx-\mu\epsilon_0$, where $k_F=p_F/\hbar$
and $\epsilon_0=\Delta_0/(k_F\xi).$
 For conventional superconductors, the interlevel spacing
 $\epsilon_0$ is much less than the superconducting gap $\Delta_0$ since
$(k_F\xi)\gg 1,$ therefore the CdGM spectrum can be considered
continuous as a function of the impact parameter of quasiclassical
trajectory $b=-\mu/k_F.$ The presence of the QP states bounded in
the vortex core was confirmed in scanning tunnel spectroscopy
(STS) experiments by the observation of the zero-bias conductance
peak at the vortex core\cite{STS}. The analogous effect of the
zero-bias conductance enhancement due to the resonant tunneling
into the midgap surface states was thoroughly studied in high
temperature D-wave superconductors\cite{Tanaka}.

 The goal of our work is to develop a theory to
calculate the zero-bias conductance of N/S junction in case when
the external magnetic field generates vortices in the S region
near the N/S interface. We consider the charge transport across
the direction of applied magnetic field. The N and S regions are
assumed to be separated by the interfacial barrier, suppressing
the Andreev reflection and the electron transport. We predict the
strong enhancement of the Andreev reflection due to the resonant
tunneling of QP through the barrier into the CdGM states localized
within the vortex cores. The essential physics of this effect is
analogous to the one which takes place in double-barrier resonant
tunneling diode\cite{DBD}. The resonant tunneling in
double-barrier quantum well structures occurs if the energy of
incident QP wave coincides with the resonant energy, then the
reflection probability is effectively suppressed due to the
interference of the QP waves within the quantum well. In our case
the quantum well is represented by the vortex core and the bounded
low energy QP state consists of coupled electron and hole waves of
almost the same amplitude. Therefore, if the incident electron has
resonant energy, then the reflected electron wave is suppressed
and the hole wave leaking from the vortex core contributes to the
Andreev reflection probability. The important difference between
our situation and the double- barrier diode case is that the
spectrum of bound states is very dense with the characteristic
interlevel spacing $\epsilon_0\sim \Delta_0/(k_F\xi)$ being much
smaller than the bulk energy gap $\Delta_0.$ At the same time, the
broadening of this levels due to the finite barrier transparency
can be estimated as $\delta E\sim \Delta_0 T\;e^{-2a/\xi},$ where
$T$ is the transparency of the interfacial barrier and the factor
$e^{-2a/\xi}$ is due to the exponential decay of subgap QP at the
superconducting slab of the thickness $a,$ which is in fact the
distance from the vortex to the N/S interface, $\xi$ is the
superconducting coherence length. Hereafter in this paper we will
neglect the discreteness of the bound energy levels assuming that
$T\;e^{-2a/\xi}\gg (k_F\xi)^{-1}.$ In fact this condition is not
very restrictive
 since $k_F\xi$ is large in many
superconducting materials, for example $k_F\xi\sim 3\cdot 10^2$ in
Nb and $k_F\xi\sim 10^4$ in Al. Neglecting the discreteness of the
spectrum of bounded QP states we can use the quasiclassical
approximation of QP quantum mechanics (see e.g.
Ref.\onlinecite{Bardeen}). Within such approximation QP move along
linear trajectories, i.e. the straight lines along the direction
of QP momentum
${\bf{n}}={\bf{k_F}}k_F^{-1}=(\cos\theta_p,\sin\theta_p).$ Note,
that for the N/S point contacts of atomic size in magnetic field
it is necessary to take into account the non-quasiclassical
divergence of the electron and hole
trajectories\cite{Backscattering}.
In the present work we consider the transport properties of wide
N/S junction so that its transverse dimension $L_y$ is much larger
than the Fermi wave length $L_y\gg\lambda_F=2\pi/k_F,$ therefore
the trajectory divergence can be neglected. Also we assume that
$L_y$ is much larger than the distance from the vortices to the
N/S interface $a$ and the characteristic intervortex distance
$L_v$. The important point is that the total QP reflection
probabilities can be found as a sum of individual vortex
contributions. Indeed, the intervortex distance is much larger
than the Fermi wavelength since $L_v\geq \xi$ and $k_F\xi\gg 1$.
Therefore, the quasiclassical trajectories (except those with
$\theta_p=\pm\pi/2$) can pass through not more than one vortex
core. For some directions of QP momentum two resonant trajectories
(i.e. passing through the vortex cores) are coupled by the normal
reflection at the interfacial barrier. Assuming the specularly
reflecting barrier this coupling occurs for the momentum angles at
the narrow angle domains near $\theta_p=\arctan(nL_v/2a),$ where
$n$ is integer. The width of the resonant angle domains
$\delta\theta\sim Te^{-2a/\xi}\ll\pi$ is determined by the width
of the resonant vortex core levels. Therefore, the contribution of
such trajectories to the amplitude of the reflected QP waves is
negligible. Then, evaluating the amplitude of the QP wave we can
separate the resonant trajectories coupled with bounded QP states
localized within the different vortex cores. Since different
trajectories do not interfere with each other, the resulting
reflection probabilities and the conductance can be found as a sum
of contributions from individual vortices.

 The dimensionless
conductance (further we will measure it in terms of the
conductance quantum $e^2/\pi\hbar$) induced by a single vortex at
zero temperature can be estimated as follows: $G_v\sim N_r
e^{-2a/\xi}T.$ Here $N_r\sim k_F\xi$ is the number of transverse
modes of the N/S junction which effectively interact with the QP
states bounded within the vortex core
 ($\xi$ is the characteristic transverse size of
the vortex core). The factor $e^{-2a/\xi}T$ is the one-particle
tunneling probability through the barrier consisting of the
insulating layer at the N/S interface and the superconducting slab
of the thickness $a$. The total vortex-induced conductance is the
sum of the individual vortex contributions $G_{vt}=n_v G_v,$ where
$n_v=L_y/L_v$ is the total number of vortices near the N/S
interface, $L_v$ is the intervortex spacing and $L_y$ is the
transverse size of the junction. The resonant mechanism of Andreev
reflection exists along with the usual non-resonant scheme
involving the two-particle tunneling through the interfacial
barrier with the probability $T^2.$ The corresponding zero-bias
conductance can be estimated as $G_0\sim N_0 T^2,$ where
$N_0=k_FL_y/\pi$ is the total number of transverse modes in N/S
junction. Therefore, the total conductance of the N/S junction in
magnetic field can be evaluated as follows:
 \begin{equation}\label{answer0}
  G=\alpha N_0T^{2}+ n_v\beta N_r e^{-2a/\xi}T,
 \end{equation}
where the coefficients $\alpha, \beta\sim 1$.  Then, for the
strong barrier $T\ll 1,$ the conductance induced by the vortex
chain with spacing $L_v$ becomes dominant at $a<a_c,$ where the
threshold distance $a_c\sim (\xi/2)\ln (\xi/TL_v)$ can be much
larger than the vortex core size and the coherence length.

The parameters of the vortex lattice can be estimated as follows:
$a, L_v\sim \sqrt{\phi_0/B},$ where $B$ is the average magnetic
field of the superconducting sample and $\phi_0=\pi\hbar
 c/e$ is the flux quantum. Then, the magnetic filed dependence of the
 vortex-induced second term in Eq.(\ref{answer0}) conductance is given by:
\begin{equation}\label{answer}
  G_{vt}\sim N_0\sqrt{\frac{B}{H_{c2}}}e^{-2\sqrt{H_{c2}/B}}T,
\end{equation}
 where $H_{c2}\sim \phi_0/\xi^2$ is the upper critical magnetic
 field of the superconductor. Note, that at zero magnetic field
 $B=0$ the conductance is $G\sim N_0 T^2$ and at $B=H_{c2}$ the
 conductance is much larger: $G\sim N_0 T$. Therefore, we suggest
 that if the surface barrier is high $T\gg 1,$ it should be
 possible to observe in experiment the field-induced increase of the conductance
 according to the Eq.(\ref{answer}).

 In the present work we do not take into account the subgap QP states which
  exists near the surface of superconductor due to the Mejssner screening of external
  magnetic field\cite{pinkus}. Such QP levels lie higher at the energy scale than the CdGM
  states if the density of supercurrent near the surface
 is less then the critical value. Therefore, as long as the zero-bias conductance is considered
 the influence of the surface QP states can be
 neglected.

The paper is organized as follows. In Sec. \ref{1} a description
of the model and the basic equations are given. In Sec. \ref{2} we
solve the scattering problem to find the Andreev and normal
reflection probabilities. Sec. \ref{3} is devoted to the
conductance calculation and  Sec.\ref{4} to the discussion of
obtained results. Finally, conclusions are given in Sec. \ref{5}.

\section{Model and basic equations}
\label{1}
\begin{figure}[hbt]
\centerline{\includegraphics[width=1.0\linewidth]{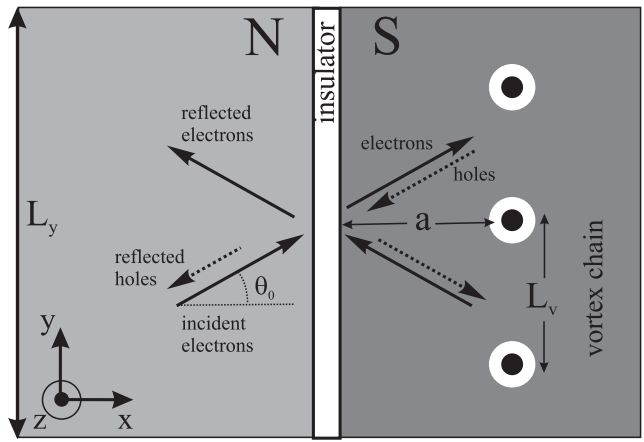}}
\caption{\label{Model} Geometry of N/S junction and sketch of QP
trajectories. The width of the junction is $L_y.$ External
magnetic field directed along $z$ axis introduces vortices in
superconductor. The distance from the first vortex chain to the
N/S interface is $a,$ and the intervortex spacing is $L_v.$ The
electrons are injected with the incident angle $\theta_0$ and
experience Andreev and normal reflection at the N/S interface.}
\end{figure}
Shown on the Fig.(\ref{Model}) is the scheme of the N/S junction
with vortex lines in the S region parallel to the N/S interface.
For the sake of simplicity we assume that there is only one
quantized QP mode in $z$ direction and take into account only the
QP motion in $xy$ plane, perpendicular to the vortex lines.

Considering one vortex from the array, the coordinate system is
chosen so that the $z$ axis coincides with the vortex line and the
origin at the $xy$ plane coincides with vortex phase singularity
point. Neglecting the suppression of the superconductivity near
the N/S interface due to the proximity effect we assume that at
  $x>-a$ (superconducting region) the order parameter can be taken as follows:
\begin{equation}\label{OP}
  \Delta({\bf{r}})=\Delta_0D_v({\bf{r}})e^{i\Phi({\bf{r}})},
\end{equation}
Here $\Delta_0$ is the gap value far from the vortex core,
$D_v({\bf{r}})$ and $\Phi({\bf{r}})$ are the dimensionless profile
and the phase of the order parameter. The particular form of
$D_v({\bf{r}})$ is not essential for our consideration, therefore
it can be chosen similar to the model profile of the isolated
vortex core\cite{Clem}: $D_v({\bf{r}})=r/\sqrt{r^2+\xi^2},$ where
$\xi$ is the coherence length. The phase distribution
$\Phi({\bf{r}})$ consists of a singular part
$\Phi_v({\bf{r}})=arg({\bf{r}})$ and a regular part
$\Phi_r({\bf{r}}),$ determined by the particular metastable vortex
lattice configuration realizing near the boundary.

  In principle, the method developed in the present paper
  is applicable to the arbitrary order parameter phase distribution
  corresponding to the metastable vortex configuration. At first, we solve a
  generic problem of the influence of a single vortex near the N/S surface on the
  zero-bias conductance of the junction.
  The vortex stability condition given by the London model,
 requires vanishing regular part of the superfluid velocity at the vortex position:
  $(\nabla \Phi_r-(2\pi/\phi_0){\bf{A}})({\bf{r}}=0)=0,$ where
  ${\bf{A}}$ is the vector potential. On the microscopic level,
  this condition is necessary for the existence of the CdGM QP
  states forming the vortex phase singularity\cite{Volovik-1993,Gygi}.
  The next step is the summation of the individual vortex
  contributions to the conductance which add independently.
  The particular vortex configuration near the boundary depends on
  many factors, such as the random pinning potential, geometry of
  the superconducting sample, magnetization history, etc.
  To estimate the dependence of the conductance on the magnetic field we
  consider the model situation assuming that the vortices near
   the boundary of superconductor sit periodically on chain with an intervortex spacing $L_v$
   at a distance $a$ from the N/S interface. We take $a$ and $L_v$ as external parameters
 of the order $a, L_v\sim\sqrt{\phi_0/B},$ where $B$ is the average magnetic field of the superconducting sample.
 The influence of the next
 vortex chains on the conductance can be neglected due to the
 rapid decay of the QP tunneling probability with the growing distance from vortices to the N/S
 interface.

 The expression for
the dimensionless zero- bias conductance of the N/S junction can
be written as follows(see Ref.\onlinecite{BKT}):
\begin{equation}\label{cond1}
  G= \frac{N_0}{2}\int_{-\pi/2}^{\pi/2}\left(1-R_n(\theta_0)+R_a(\theta_0)\right)\cos\theta_0\,d\theta_0,
\end{equation}
where $R_n(\theta_0)$ and $R_a(\theta_0)$ are the probabilities of
normal and Andreev reflection respectively, $\theta_0$ is the
incident angle: ${\bf{k_F}}=k_F(\cos\theta_0,\sin\theta_0)$. The
total number of propagating modes $N$ is determined by the channel
width: $N_0=k_FL_y/\pi.$
 The problem of QP scattering at the N/S interface is
 formulated within the BdG theory. The equation for the
 electron and hole waves coupled by the superconducting gap
$\Delta({\bf{r}})$ reads as follows:

\begin{eqnarray}
 \label{BdG} \left( \hat H_0({\bf{r}})\;\;\;\;\; \Delta({\bf{r}})\atop
\Delta^*({\bf{r}})\;\;\;\; -\hat
H_0({\bf{r}})\right)\hat\Psi({\bf{r}})=E\hat\Psi({\bf{r}}),
    \end{eqnarray}
    Here
$$\hat H_0({\bf{r}})=\frac{1}{2m}\left(\hat
{\bf{p}}-\frac{e}{c}{\bf{A}}\right)^2-E_F+V(x)$$
 with $\hat
{\bf{p}}=-i\hbar\nabla,$ $\hat\Psi({\bf{r}})=(u({\bf{r}}),
v({\bf{r}})).$
  The interfacial barrier separating the N and S regions is modeled by the repulsive
  delta potential $V(x)=H\delta(x),$
  parameterized by the dimensionless barrier strength $Z=H/\hbar V_F$\cite{BKT}.
    The boundary conditions at the $N/S$ interface are:
 \begin{equation}\label{bc}
  \left[\hat\Psi(-a)\right]=0,
 \end{equation}
 \begin{equation}\label{bc1}
 \left[\partial_x\hat\Psi(-a)\right]=(2k_FZ)\hat\Psi(-a),
 \end{equation}
  where  $[f(x)]=f(x+0)-f(x-0)$.

 To overcome the complexity of the scattering problem coming from the
broken spatial invariance of the superconducting gap, we treat the
Eq.(\ref{BdG}) within quasiclassical approximation.
Generally, the quasiclassical form of the wave function can be
constructed as follows: $\hat
\Psi({\bf{r}})=e^{i{\bf{k_F}}\cdot{\bf{r}}}\hat f({\bf{r}})$,
where $\hat f({\bf{r}})=(U({\bf{r}}),V({\bf{r}}))$ is a slow
varying envelope function. Then the system (\ref{BdG}) reduces to
the system of the first-order quasiclassical equations along the
linear trajectories defined by the direction of the QP momentum
${\bf{n}}={\bf{k_F}}k_F^{-1}=(\cos\theta_p,\sin\theta_p).$ Each
trajectory is specified
  by the angle $\theta_p$ and the impact parameter
$b=r\sin(\theta-\theta_p),$ where $\theta$ is the polar angle:
${\bf r}=-r(\cos\theta,\sin\theta).$ Introducing the coordinate
along trajectory
$s=({\bf{n}}\cdot{\bf{r}})=-r\cos(\theta_p-\theta)$ we arrive at
the following form of the quasiclassical equation: $\hat H\hat
f=\epsilon\hat f,$ with the hamiltonian:

\begin{equation}\label{quasiclass}
  \hat
  H=-i\xi\hat\sigma_z\partial_s+F+D_v\left(\hat\sigma_x\cos\Phi-\hat\sigma_y\sin\Phi
  \right),
\end{equation}
 where $\epsilon=E/\Delta_0,$ $\xi=\hbar V_F/\Delta_0$ is the coherence length,
 $D_v({\bf{r}})$ and $\Phi({\bf{r}})$ are the dimensionless magnitude
 and phase of the order parameter
 and $F({\bf{r}})=(\pi\xi/\phi_0){\bf{n}}\cdot{\bf{A}}.$

 Considering
 the zero-bias problem we will have to analyze only the zero-energy
excitations with $\epsilon=0$. Within the normal metal region at
$x<-a,$ neglecting the influence of magnetic field on the QP
motion, the quasiclassical equation (\ref{quasiclass}) becomes
trivial: $\partial_s \hat f(s,b)=0.$ It means that the envelope
function is constant along the trajectories.

Obviously, this is not the case at the superconducting region
$x>-a$, where the electron and hole waves are
  coupled.  Note that for the wave functions at
  the S region corresponding to the zero energy the following
  representation can be used\cite{Bardeen}: $\hat f =e^{\zeta}\left(e^{i(\eta+\Phi)/2},
  e^{-i(\eta+\Phi)/2}\right)$, where $\zeta =\zeta (s,b)$
  and $\eta =\eta (s,b)$ are real-valued functions. Then, the quasiclassical
  equation (\ref{quasiclass})
  can be written as follows:

  \begin{equation}\label{eta}
\xi\partial_s\eta+2D_v\cos\eta+\epsilon_d=0,
 \end{equation}

 \begin{equation}\label{dzita}
\xi\partial_s\zeta+2D_v\sin\eta=0.
\end{equation}
where $\epsilon_d({\bf{r}})=\hbar{\bf{k}_F}{\bf{v_s}}/\Delta_0$ is
the dimensionless Doppler shift energy. For the wave functions
$\hat f_\pm $ decaying at the different ends of the trajectory
$\hat f_\pm(s=\pm\infty)=0$ from Eq.(\ref{dzita}) we obtain:
\begin{equation}\label{tetaBC}
\eta_\pm(s=\pm\infty)=\pm\pi/2.
\end{equation}

As we will see below, the main contribution to the enhanced
Andreev reflection probability comes from the trajectories with
small impact parameters $|b|\ll\xi,$ passing through the vortex
core. For such trajectories neglecting the nonsingular part of the
superfluid velocity near the vortex core the analytical solution
of Eq.(\ref{eta}) can be obtained following the results of
Ref.\onlinecite{MelnikovHeat}:

\begin{equation}\label{etaPM}
  \tan\eta_\pm=0.5\left(A^{-1}_\pm e^{-2K(s)}-A_\pm e^{2K(s)}\right),
\end{equation}
$A_\pm  =\gamma (b)(sgn(s)\mp 1),$ where $\gamma (b)=-\omega b,$

 $$
 K(s)=\frac{1}{\xi}\int\limits_0^s D_v(s^\prime)
 ds^\prime=\sqrt{(s/\xi)^2+1}-1,
  $$
 $$
\omega=\frac{1}{\xi}\int_0^{\infty}\frac{D_v(s)}{s} e^{-2K(s)}ds.
$$

    \section{Scattering problem: normal and Andreev reflection probabilities}
\label{2}
 The boundary conditions (\ref{bc},\ref{bc1}) determine the specularly reflecting N/S interface, coupling the waves with
 wave vectors ${\bf{k}}_F=k_F(\cos\theta_0,\sin\theta_0),$
  and ${\bf{k^\prime}}_F=k_F(\cos(\pi-\theta_0),\sin(\pi-\theta_0))$. Therefore
 if the incident electron wave is $u_i=e^{i{\bf{k_F}}{\bf{r}}},$
 then reflected electron $u_r$ and hole $v_r$ waves will have the form
 $$u_r=U_re^{i{\bf{k^\prime_F}}{\bf{r}}},\;\;v_r=V_re^{i{\bf{k_F}}{\bf{r}}},
$$
 where and $U_r(b,s)$ and $V_r(s,b)$ are the envelope
 functions. Each point $(-a,y)$ at the N/S interface
 lies on the intersection of two quasiclassical trajectories, characterized by the
 angles $\theta_p=\theta_0$ and $\theta_p=\pi-\theta_0.$
   From the simple trigonometry it is easy to see that
  the impact parameters of these trajectories are $b_+ =-a\sin(\theta_0-\theta)/\cos\theta$ and $b_- =-a\sin(\theta_0+\theta)/\cos\theta$
   correspondingly, where $\theta=-\arctan(y/a)$ is the polar
  angle. The coordinate of the intersection point is
  $s_+=-a\cos(\theta-\theta_0)/\cos\theta$
 and $s_-=a\cos(\theta+\theta_0)/\cos\theta$ for the trajectories
 characterized by the angles $\theta_p=\theta_0$ and $\theta_p=\pi-\theta_0$
 correspondingly. Then, the reflection probabilities are given by:
 \begin{equation}\label{Rn}
  R_n(\theta_0)=\frac{a}{L_y}\int^{\alpha}_{-\alpha}|U_r(\theta,\theta_0)|^2(\cos\theta)^{-2}d\theta
 \end{equation}
\begin{equation}\label{Ra}
  R_a(\theta_0)=\frac{a}{L_y}\int^{\alpha}_{-\alpha}|V_r(\theta,\theta_0)|^2(\cos\theta)^{-2}d\theta,
  \end{equation}
where the integration is done over the N/S interface,
$\alpha=\arctan (L_y/2a),$ $U_r(\theta,\theta_0)=U_r(b_-,s_-)$
 and
 $V_r(\theta,\theta_0)=V_r(b_+,s_+).$

  Following the usual procedure,
to find the reflected wave amplitudes $U_r(\theta,\theta_0)$ and
$V_r(\theta,\theta_0)$ one needs to match the N and S regions
solutions at the N/S interface.
   For the envelope functions the boundary conditions
  (\ref{bc},\ref{bc1}) yield:

 $$
  1+ U_r=
   e^{i\eta_+/2}C^++e^{i\eta_-/2}C^-,
 $$
 $$
 V_r= e^{-i\eta_+/2} C^++ e^{-i\eta_-/2}C^-,
 $$
 $$
 \left(1- U_r\right)+2iZ\left(1+ U_r\right)=
  e^{i\eta_+/2}C^+-e^{i\eta_-/2}C^-,
 $$
 $$
 V_r(1+2iZ)= e^{-i\eta_+/2} C^+- e^{-i\eta_-/2}C^-,
 $$
$C^+,C^-$ are arbitrary
 constants and $\eta_\pm=\eta_\pm(s_\pm,b_\pm),$ where $\eta_\pm(s,b)$ are
 the solutions of Eq.(\ref{eta}) with boundary conditions
 (\ref{tetaBC}) along the trajectories with $\theta_p=\theta_0$ and $\theta_p=\pi-\theta_0$ for the upper and lower signs correspondingly.

  Solving this system we obtain:
 \begin{equation}\label{Ur}
 U_r(\theta,\theta_0)=-\frac{(1-e^{i\chi})(\tilde{Z}^2-i\tilde{Z})}{1+\tilde{Z}^2(1-e^{i\chi})},
 \end{equation}
 \begin{equation}\label{Vr}
  V_r(\theta,\theta_0)=\frac{e^{-i\eta_+}}{1+\tilde{Z}^2(1-e^{i\chi})},
 \end{equation}
  where $\chi(\theta,\theta_0)=\eta_--\eta_+$ and $\tilde{Z}=Z/\cos\theta_0.$

 Note, that for the small impact parameters $|b_\pm|\ll\xi$ the factors
 $e^{i\eta_\pm}$ can be obtained analytically with the help
 of Eq.(\ref{etaPM}) as follows:
 \begin{equation}\label{eta+}
 e^{i\eta_+}=i\frac{J-i(\theta-\theta_0)}{J+i(\theta-\theta_0)},
 \end{equation}

 \begin{equation}\label{eta-}
 e^{i\eta_-}=-i\frac{J-i(\theta+\theta_0)}{J+i(\theta+\theta_0)},
 \end{equation}
where $J=e^{-2K(a/\cos\theta)}\cos\theta/(2\omega a).$ For the
small angles $|\theta|,|\theta_0|\ll \xi/a$ the equations
(\ref{eta+}) and (\ref{eta-}) are valid simultaneously yielding:
 \begin{equation}\label{hi}
  e^{i\chi}=-\frac{J^2+\theta^2-\theta_0^2-2i\theta_0J}
  {J^2+\theta^2-\theta_0^2+2i\theta_0J}.
\end{equation}

 \section{Vortex-induced zero-bias conductance}
\label{3}

Now, using the expressions (\ref{Ur}, \ref{Vr}) for the amplitudes
of reflected waves and reflection probabilities (\ref{Rn},
\ref{Ra}) it is possible to find the zero-bias conductance.
Introducing the function
$g(\theta,\theta_0)=1-|U_r(\theta_0,\theta)|^2+|V_r(\theta_0,\theta)|^2$
the expression for the dimensionless conductance (\ref{cond1})
reads as follows:

\begin{equation}\label{cond0}
  G=\frac{k_Fa}{2\pi}\int^{\alpha}_{-\alpha}
 (\cos\theta)^{-2} d\theta \int_{-\pi/2}^{\pi/2}
 g(\theta,\theta_0)\cos\theta_0\,d\theta_0,
\end{equation}
 where $\alpha=\arctan(L_y/2a).$
 It is convenient also to introduce here a local conductivity, i.e. the conductance per unit length of the N/S surface:
  $$\sigma(\theta)=\frac{k_F}{2\pi}\int_{-\pi/2}^{\pi/2}
 g(\theta,\theta_0)\cos\theta_0\,d\theta_0.$$
 Employing Eqs.(\ref{Ur},\ref{Vr})
we obtain:
 \begin{equation}\label{cond2}
  g(\theta,\theta_0)=\frac{2}{(\tilde{Z}^4+\tilde{Z}^2)|1-e^{i\chi}|^2+1}.
\end{equation}

 If the applied magnetic field is zero and the superconductor is homogeneous $\chi=\pi,$
 we obtain $g(\theta,\theta_0)=g_0(\theta_0),$ where
 $g_0(\theta_0)=(1/2)(\tilde{Z}^2+1/2)^{-2}.$ Then, the vortex-induced part of
 the conductivity is given by
$$\sigma_{v}(\theta)=\frac{k_F}{2\pi}\int_{-\pi/2}^{\pi/2}
 g_v(\theta,\theta_0)\cos\theta_0\,d\theta_0,$$
 where $g_v=g-g_0:$
  $$
 g_v=\frac{(\tilde{Z}^4+\tilde{Z}^2)\left(4-|1-e^{i\chi}|^2\right)}{2(\tilde{Z}^2+1/2)^2
 \left((\tilde{Z}^4+\tilde{Z}^2)|1-e^{i\chi}|^2+1\right)}.
 $$

 To start the analysis of Eq.(\ref{cond2}) let's note that for the low surface barrier $\tilde{Z}\rightarrow 0$
 we get $g_v(\theta,\theta_0)=0.$ In this case all the incident QP undergo Andreev reflection and the zero-bias conductance
 is the same as in case of homogeneous superconductor: $G=2N_0$.
  As the barrier becomes higher, the Andreev reflection is suppressed and the conductance is reduced.

  The function $g_v(\theta,\theta_0)$
   reaches its maximum
   $$g_{vm}=2\frac{\tilde{Z}^4+\tilde{Z}^2}{(\tilde{Z}^2+1/2)^2}$$ if $|1-e^{i\chi}|=0.$ In fact this condition
   determines the resonant trajectories, corresponding to the zero-energy vortex core states modified by the
   normal reflection from the interfacial barrier. The resonant trajectories should pass through the vortex core
   therefore the function $g_v(\theta,\theta_0)$
   has a sharp peak at $\theta_0\approx\pm\theta.$ The
   width of this peak is determined by the barrier strength and the distance from the vortex to the surface.
    For the small angles
   $|\theta|,|\theta_0|\ll \xi/a$ with the help of
   Eq.(\ref{hi}) we obtain:

  \begin{equation}\label{gv}
  g_v(\theta_0,\theta)=\frac{g_{vm}J_0^2\theta_0^2}{(\tilde{Z}^2+1/2)^2(\theta_0^2-\theta^2-J_0^2)^2+J_0^2\theta_0^2},
\end{equation}
where $J_0=e^{-2K(a)}/(2\omega a)\sim (\xi/a)e^{-2a/\xi},$ which
is a small parameter since $J\ll 1$ for $a\geq\xi.$ The maximum of
$g_v(\theta,\theta_0)$ determined by Eq.(\ref{gv}) lies at
$\theta_0^2=\theta^2+J_0^2$.

 Employing Eq.(\ref{gv}) it is easy to compute the vortex-induced part of the
 conductivity $\sigma_{v}(\theta)$ at the small angle domain $|\theta|\ll\xi/a.$ The main
contribution to integral over $\theta_0$ comes
 from the small vicinity of the point $\theta_0=\theta$. Then with good accuracy we
 obtain: $\sigma_{v}=\sigma_{v0},$ where
 \begin{equation}\label{g1}
 \sigma_{v0}=k_F J_0\frac{Z^4+Z^2}{(Z^2+1/2)^3}.
 \end{equation}

\begin{figure}[hbt]
\centerline{\includegraphics[width=0.6\linewidth]{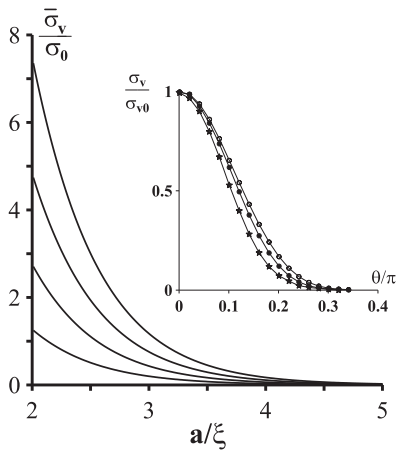}}
\caption{\label{F(t)} Plot of the ratio $\bar{\sigma}_v/\sigma_0$
of the average vortex induced conductivity to the conductivity in
the absence of vortices. Curves from top to bottom correspond to
$Z=5,4,3,2$. Inset: function $\sigma_{v}(\theta)/\sigma_{v0}$ for
$a/\xi=2$ (open circles), $a/\xi=3$ (filled circles) $a/\xi=5$
(asteriskes); $Z=2.$}
\end{figure}

  At larger angles the function $\sigma_v(\theta)$ can be evaluated
only numerically.  Numerical calculation described below shows
that $\sigma_{v}(\theta)$ is maximal at $\theta=0$ and steadily
decreases to zero as $|\theta|\rightarrow \pi/2.$
  (see inset on Fig.\ref{F(t)}). Then, the resulting conductance induced by a single vortex
$G_v=a\int^{\alpha}_{-\alpha}
 (\cos\theta)^{-2} \sigma_v(\theta)d\theta$
  is given by:
 \begin{equation}\label{VIcond}
  G_{v}=\beta ( k_F
\xi)e^{-2K(a)}\frac{Z^4+Z^2}{(Z^2+1/2)^3},
 \end{equation}
    where $\beta=(2\omega\xi)^{-1}\int^{\alpha}_{-\alpha}
 (\cos\theta)^{-2}\sigma_v(\theta)/\sigma_{v0} d\theta\sim 1.$

 To evaluate the conductance rigorously, we find the factor
 $e^{i\chi}$ in Eq.(\ref{cond2}) and then the reflection probabilities solving numerically Eq.(\ref{eta})
 with boundary conditions (\ref{tetaBC}).
  We assume that the
     regular part of the phase distribution is $\Phi_r({\bf{r}})=-arg({\bf{r}}-{\bf{r_{av}}})$
   corresponding to the image vortex situated at the point ${\bf{r_{av}}}=(-2a,0,0)$
   behind the N/S interface.
  The vector potential is chosen as ${\bf{A}}=B\left[{\bf{z_0}}\times({\bf{r}}-{\bf{r_0}})\right]/2,$
   where ${\bf{r_{0}}}=(-a,0,0)$ is the point at the boundary between
   the vortex and the image vortex, and ${\bf{z_0}}$ is the unit vector along the $z$ axis. Within such model the
   condition of vanishing current through the N/S interface $(\partial_x \Phi-(2\pi/\phi_0){\bf{A}}_x)=0$
    is satisfied automatically, and the
   vortex stability is achieved by setting $a=\sqrt{\phi_0/B}.$
   The numerical plot of the function $\sigma_{v}(\theta)/\sigma_{v0}$ at different
   distances $a$ from the vortex to the interface is presented on the inset at Fig.(\ref{F(t)}). The maximum
   value $\sigma_{v}(\theta=0)$ with good accuracy coincides with the analytical
   estimation given by (\ref{g1}). The coefficient $\beta$ in Eq.(\ref{VIcond}) is found to be nearly
   constant as a function of $a$: it decreases slightly from $\beta\simeq 0.6$ at $a=2\xi$ to $\beta\simeq 0.4$ at $a=5\xi.$
   On the Fig.(\ref{F(t)}) for the various magnitude of barrier strength we plot the ratio $\bar{\sigma}_v/\sigma_0$
   of the average vortex-induced conductivity $\bar{\sigma}_v=G_v/\xi$ to the
   conductivity of the N/S junction in the
 absence of vortices given by: $$\sigma_0=\frac{k_F}{2\pi}\int_{-\pi/2}^{\pi/2}
 g_0(\theta,\theta_0)\cos\theta_0\,d\theta_0.$$

\section{Discussion}\label{4}

\begin{figure}[hbt]
\centerline{\includegraphics[width=1.0\linewidth]{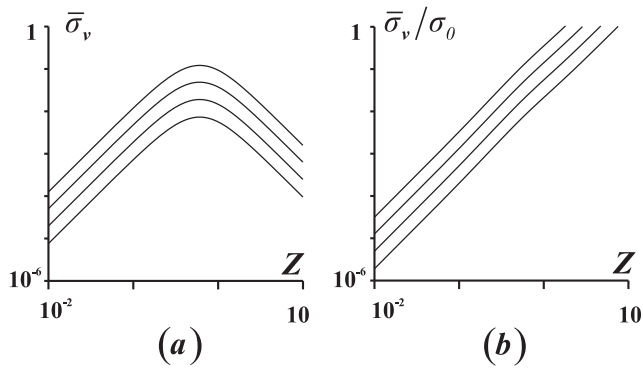}}
\caption{\label{Gv(T)} (a): Plot of the average vortex-induced
conductivity $\bar{\sigma}_v$ as a function of the  barrier
strength $Z$ in logarithmic scale.  (b): Plot of the ratio
$\bar{\sigma}_v/\sigma_0$ as a function of the barrier strength
$Z$ in logarithmic scale. Curves from top to bottom correspond to
$a/\xi=2,\;2.5,\;3,\;3.5$.}
\end{figure}

Considering the strong barriers $Z\gg 1$ Eq.(\ref{VIcond}) can be
written as $G_{v}=\beta ( k_F \xi)e^{-2K(a)}T,$ where we
introduced the barrier transparency $T=(1+Z^{2})^{-1}\approx
Z^{-2}$. A simple understanding of this result can be obtained
within the framework of the tunneling hamiltonian
approach\cite{Mahan}. The conventional expression for the
tunneling conductivity of the N/S junction at zero temperature
reads:
\begin{equation}\label{tunn_hom}
  \sigma=\sigma_n\nu /\nu_0,
\end{equation}
 where $\nu$ is a
 local superconducting DOS at the Fermi level and $\sigma_n$ is a normal state
 tunneling conductivity.   For the S-wave superconductor the transformation of vortex core
states near the surface can be neglected at first
approximation\cite{Schopohl}. Then the local DOS near the surface
at the Fermi level is determined by the density of vortex core
states $\nu_v({\bf{r}})$ given by\cite{Maki, Volovik1}:
\begin{equation}\label{DOS}
  \nu_v({\bf{r}})=\frac{1}{2\pi}\int_0^{2\pi}
  |\hat f({\bf{r}},\theta_p)|^2\delta(\epsilon_0k_Fr\sin(\theta-\theta_p))d\theta_p,
\end{equation}
 where $\hat f({\bf{r}},\theta_p)$ is the envelope of
 the QP wave function. For the CdGM wave functions $|\hat f({\bf{r}},\theta_p)|^2\sim e^{-2K(r)}k_F/\xi,$
 then $\nu_v({\bf{r}})\sim \nu_0(\xi/r)e^{-2K(r)},$ where $\nu_0=m/\hbar^2$ is the two-dimensional DOS at the normal metal.
   Substituting $\nu=\nu_v(r=a/\cos\theta)$ and $\sigma_n\sim Tk_F$ we
 obtain the conductivity: $\sigma\sim k_FTJ(\theta),$ which
 coincides to the order of magnitude with expression (\ref{g1}) if $Z\gg 1$ and $|\theta|\ll 1.$
   Integrating
 $\sigma(\theta)$ over the N/S interface we arrive at
  expression (\ref{VIcond}) for the conductance with
factor $\beta$
 given by:  $\beta\sim\int_{-\alpha}^{\alpha}
 \exp(2K(a)-2K(a/\cos\theta))(\cos\theta)^{-1} d\theta.$ Note, that although yielding the qualitatively right
 answer for the vortrex-induced conductance, the tunneling hamiltonian approach drops out the
 contribution from the non-resonant Andreev reflection with the probability of the order $T^2.$ As we will see below,
 the contributions of the resonant and non-resonant Andreev reflections can be comparable even
 when the surface barrier is rather strong.
  Certainly, the tunneling hamiltonian approach fails to provide
 the answer if the barrier strength is not very high, when
  the influence of vortices on the conductance
 is reduced and the non-resonant Andreev reflection prevails.
  On the Fig.\ref{Gv(T)}a we show in logarithmic scale the vortex induced conductivity
$\bar{\sigma}_v$ as a function of
 the barrier strength $Z$ for the several values of the distance
 $a$. At small values of $Z$ the function $\bar{\sigma}_v (Z)$
 grows $\bar{\sigma}_v \sim Z^2$ in accordance with the estimation
 (\ref{g1}). At larger values of the barrier strength $Z\gg 1$ the behaviour of
 $\bar{\sigma}_v (Z)$ changes to $\bar{\sigma}_v \sim Z^{-2}$. But
 at the same time the conductivity without vortices at $Z\gg 1$ behaves as $\sigma_0\sim
 Z^{-4}$. Therefore, the ratio $\bar{\sigma}_v/\sigma_0$ is monotonically growing as a function of
 the barrier strength $Z$ proportional to $Z^2$ (see Fig.\ref{Gv(T)}b).

  Let's have a look at the expression for the total conductance of the N/S junction, which has quite a simple form if $Z\gg 1.$
  Neglecting edge effects end summing up the individual vortex
  contributions we obtain:
    \begin{equation}\label{cond4}
  G=(8/15)N_0T^{2}+ n_v\beta(k_F\xi)e^{-2K(a)}T,
 \end{equation}
where $n_v=L_y/L_v$ is the total number of vortices near the N/S
interface. The obtained expression for the total conductance
(\ref{cond4}) consists of two terms. The
 first term $G_0\sim N_0T^{2}$ coincides with the conductance of
 the N/S junction at zero magnetic field. The factor $T^2$ is determined by the probability of the sequential
 tunneling of the incident and reflected QP through the high interfacial
 barrier. The second term is the total vortex-induced conductance  $G_{vt}=n_vG_v\sim n_v (k_F\xi)e^{-2K(a)}T$;
 it comes from the tunneling of the incident
  QP into the zero energy CdGM states inside the vortex core.
 Indeed it is easy to see that $e^{-2K(a)}T,$ where
 $K(a)=\sqrt{(a/\xi)^2+1}-1$,
 is the one- particle tunneling probability through the interfacial barrier and the superconducting
 layer of the thickness $a$ with slightly suppressed gap due to the presence of the vortex.
 The factor $k_F\xi$ is the number of resonant transverse
 modes for a single vortex. The vortex-induced conductance $G_{vt}$ prevails over $G_0$ when $a<a_c$, where the critical distance $a_c$
 is determined by $a_c\approx (\xi/2)\ln (L_v/T\xi).$
  The parameters
 of the vortex configuration, such as the intervortex spacing $L_v$ and
 the distance $a$ from the vortex array to the boundary of
 superconductor are determined by the magnetic field, therefore the conductance of the N/S junction
 can be controlled by the magnetic field. Using Eq.(\ref{answer}) we
 obtain that the critical magnetic field $H_c$ when $G_{vt}\sim
 G_0$ is determined by the following transcendental equation $\ln(x/T)=2/x,$ where $x=\sqrt{B/H_{c2}}.$
  Taking for example the barrier strength $Z=5,$ we obtain that the critical field is $B_c\sim 0.5 H_{c2},$
  and the critical distance $a_c\sim 1.5\xi.$  Therefore, the influence of the
 resonant vortex core states on the conductance can become
 significant when the magnetic field is less then the upper critical
 and vortices are quite far from the N/S interface.

 Finally, we should note that in real N/S junctions the motion of
 QP is certainly affected by impurity scattering.
  The influence of
impurities can be neglected completely assuming that the life time
of the vortex core states due to the finite barrier transparency
is much shorter than the relaxation time of the QP momentum:
$\hbar/\delta E\ll\tau,$ or
 \begin{equation}\label{impurity}
  l_e\gg (Te^{-2a/\xi})^{-1}\xi,
 \end{equation}
where $l_e=V_F\tau$ is an elastic mean free path of QP at the S
region. This condition certainly can be fulfilled if the barrier
transparency is not very high, i.e. $T\sim 1$ and the vortex chain
is situated not far from the N/S interface, so that the factor
$Te^{-2a/\xi}$ is not very small. Otherwise, if the condition
(\ref{impurity}) is not fulfilled, the impurity scattering will
modify the conductance. The simplest approach to estimate the
conductance in this situation is based on the tunneling
hamiltonian, yielding the Eq.(\ref{tunn_hom}) for the
conductivity. Due to the impurity scattering the local
superconducting DOS differs from that given by the Eq.(\ref{DOS}),
which is valid for the clean case $l_e\gg\xi$. In particular, the
sharp peak at $r=0$ is smeared, therefore at the center of vortex
the DOS is smaller as compared to the clean case. But at the
distances $r>\xi$ from the vortex core (e.g. at the N/S interface)
the DOS is not suppressed by the impurities even if $l_e\sim\xi$.
On the contrary, it is even larger than in the clean case due to
the smearing of the DOS peak at the center of vortex
\cite{Ichioka}. Therefore, in case of the rather high impurity
concentration on the $S$ side the dependence of the vortex
-induced conductance on the magnetic field is still described by
the Eq.(\ref{answer}). Another important point is the influence of
impurities on the non-resonant part of the conductance, i.e. the
first term in Eq.(\ref{answer0}). In particular, the interference
of QP waves reflected from the interface barrier and impurities on
the N side of the N/S junction can also result in the low-bias
conductance enhancement, known as reflectionless tunneling (see
Ref. \onlinecite{Beenakker} and references therein). In
experiments where the reflectionless tunneling effect was observed
\cite{Kastalsky} the condition $l_e>\xi$ was fulfilled. In this
case the critical value of magnetic field suppressing the
reflectionless tunneling\cite{vanWees} $H_c\sim \phi_0/(12l_e^2)$
was much less than the upper critical field of the superconductor.
In our case the applied magnetic field must be strong enough to
create the dense vortex lattice in superconductor: $B\sim
H_{c2}\gg H_c$. Therefore, under the same experimental conditions
as in Ref.\onlinecite{Kastalsky} the reflectionless tunneling
effect is absent in the range of magnetic fields that we are
interested in.

 \section{Conclusion}
 \label{5}

To summarize, we have investigated the low-energy  charge
transport in N/S junction across the direction of applied magnetic
field. We have found the strong enhancement of the zero-bias
conductance due to the resonant tunneling of the incident QP into
the subgap vortex core states. The effect is most sound if the
conventional channel of Andreev reflection is suppressed by the
high interfactial barrier. Note that usually, the vortex core
states are investigated in STS experiments, where the charge
transport is measured along the direction of magnetic field. For
the alternative to the STS methods now we can suggest the
transport measurements in planar structure with wide-area N/S
contacts. The vortex induced conductance that we have studied
depends exponentially  on the distance from the vortex chain to
the N/S interface and therefore can be effectively controlled by
the external magnetic field. Also for the possible experimental
setup one can consider the mesoscopic superconducting sample with
lateral tunneling junctions, such as in Ref.\onlinecite{Peeters}.
Since the vortex -induced conductance is proportional to the
number of vortices, the conductance vs magnetic field dependence
should reveal pronounced steps marking the switch of the total
vorticity of the sample.

 \section{Acknowledgements}
It's my pleasure to thank Dr. Alexander S. Mel'nikov for numerous
stimulating discussions and help in preparation of this paper.
Also I am grateful to Ekaterina Ezhova for help with numeric
calculations. This work was supported, in part, by Russian
Foundation for Basic Research, by Program ``Quantum Macrophysics''
of RAS, and by Russian Science Support and ``Dynasty''
Foundations.


\begin{thebibliography}{99}
\bibitem{BKT}
G.~E. Blonder, M.~Tinkham and T.~M. Klapwijk, Phys. Rev.~B {\bf
25}, 4515 (1982).

 \bibitem{Peeters}
A.~Kanda, B.~J. Baelus, F.~M. Peeters, K.~Kadowaki, and Y.~Ootuka,
Phys. Rev. Lett. {\bf 93}, 257002 (2004); B.~J. Baelus, A.~Kanda,
F.~M. Peeters, Y.~Ootuka and K.~Kadowaki, Phys. Rev. B {\bf 71},
140502(R) (2005); B.~J. Baelus, A.~Kanda, N.~Shimizu, K.~Tadano,
Y.~ Ootuka, K.~Kadowaki, and F.~M. Peeters, Phys. Rev. B {\bf 73},
024514 (2006).

\bibitem{Mesovortices}
G.~Boato, G.~Gallinaro, and C.~Rizzuto, Solid State Commun. {\bf
3}, 173 (1965); D.~S. McLachlan, Solid State Commun. {\bf 8}, 1589
(1970); V.~A. Schweigert, F.~M. Peeters, and P.~S. Deo, Phys. Rev.
Lett. {\bf 81}, 2783 (1998); A.~K. Geim, S.~V. Dubonos, J.~J.
Palacios,I.~V. Grigorieva, M.~Henini, and J.~J. Schermer, Phys.
Rev. Lett. {\bf 85}, 1528 (2000); L.~F. Chibotaru, A.~Ceulemans,
V.~Bruyndoncx, V.~V. Moshchalkov, Nature (London) {\bf 408}, 833
(2000); A.~K. Geim, S.~V. Dubonos, J.~J. Palacios, I.~V.
Grigorieva, M. Henini, and J.~J. Schermer, Phys. Rev. Lett. {\bf
85}, 1528 (2000).
\bibitem{Tinkham} M. Tinkham, {\it Introduction to
Superconductivity} (McGraw-Hill, New York, 1996), 2nd ed., Chap.
10.

\bibitem{esteve}
A.~Anthore, H.~Pothier, and D.~Esteve, Phys. Rev. Lett. {\bf 90},
127001 (2003).

\bibitem{CdGM}
C.~Caroli, P.~G. de Gennes, J.~Matricon, Phys. Lett. {\bf 9}, 307 (1964).

\bibitem{STS}
  H.~F. Hess, R.~B. Robinson, R.~C. Dynes, J.~M. Valles, Jr., and J.~V. Waszczak, Phys. Rev. Lett. {\bf 62}, 214
 (1989);H.~F. Hess, R.~B.Robinson, and J.~V. Waszczak, Phys. Rev. Lett. {\bf 64}, 2711
 (1990);A.~Kohen, Th.~Proslier, T.~Cren, Y.~Noat, W.~Sacks, H.~Berger,
 and D.~Roditchev, Phys. Rev. Lett. {\bf 97}, 027001 (2006).

\bibitem{Tanaka}
C.~R. Hu, Phys. Rev. Lett. {\bf 72}, 1526
 (1994); J.~Yang, C.~R. Hu, Phys. Rev. B {\bf 50}, 16766
 (1994); Y.~Tanaka, S.~Kashiwaya, Phys. Rev. Lett. {\bf 74}, 3451
 (1995); S.~Kashiwaya, Y.~Tanaka, M.~Koyanagi, H.~Takashima and K.~Kajimura, Phys. Rev. B {\bf 51}, 1350
 (1995).

 \bibitem{DBD}
 B.~Ricco and M.~Ya.Azbel, Phys. Rev. B {\bf 29}, 1970 (1984)

\bibitem{Bardeen}
J.~Bardeen, R.~Kummel, A.~E. Jacobs and L.~Tewordt, Phys. Rev.
{\bf 187}, 556 (1969)

\bibitem{Backscattering}
N.~B. Kopnin, A.~S. Melnikov and V.~M. Vinokur, Phys. Rev. B {\bf
71}, 052505 (2005);

 \bibitem{pinkus}
P.~Pincus, Phys. Rev. {\bf 158}, 346
 (1967);

\bibitem{Clem}
 J.~R. Clem, J. Low Temp. Phys. {\bf 18}, 427 (1975)

\bibitem{Volovik-1993}
G.~E. Volovik, Pis'ma Zh. Eksp. Teor. Fiz. {\bf 58}, 444 (1993)
[JETP Lett. {\bf 58}, 455 (1993)].

\bibitem{Gygi}
F.~Gygi and M.~Schluter, Phys. Rev. B {\bf 43}, 7609
 (1991);

\bibitem{MelnikovHeat}
A.~S. Mel'nikov, N.~B. Kopnin and V.~M. Vinokur, Phys. Rev.~B {\bf
68}, 054528 (2003)

%

\bibitem{Mahan}
G.~D. Mahan, {\it Many-particle physics} (Plenum Press, New York,
1993), 2nd ed., Chap. 9.

\bibitem{Schopohl}
S.~Graser, C.~Iniotakis, T.~Dahm and N.~Schopohl, Phys. Rev. Lett.
{\bf 93}, 247001
 (2004);

 \bibitem{Maki}
N.~Schopohl and K.~Maki, Phys. Rev. B {\bf 52}, 490
 (1995);

 \bibitem{Volovik1}
N.~B. Kopnin and G.~E. Volovik, JETP Lett. {\bf 64}, 690 (1996)

\bibitem{Ichioka}
P.~Miranovic, M.~Ichioka and K. Machida, Phys. Rev. B {\bf 70},
104510 (2004);

\bibitem{Beenakker}
C.W.J.~Beenakker, Rev. Mod. Phys {\bf 69}, 731 (1997);

\bibitem{vanWees}
B. J. van Wees, P. de Vries, P. Magnee, and T. M. Klapwijk, Phys.
Rev. Lett. {\bf 69}, 510 (1992);

\bibitem{Kastalsky}
A. Kastalsky, A. W. Kleinsasser, L. H. Greene, R. Bhat, F. P.
Milliken, and J. P. Harbison, Phys. Rev. Lett. {\bf 67}, 3026
(1991)

\end{thebibliography}
\end{document}